\documentclass[showpacs,amssymb,aps]{revtex4}
\usepackage{amsmath}
\usepackage{amstext}
\usepackage{amsopn}
\usepackage{amsfonts}
\usepackage{amssymb}
\usepackage{bbm}
\usepackage{accents}
\usepackage{empheq}
\usepackage{graphicx}
\usepackage{epsf}
\usepackage{graphics}
\usepackage[latin1]{inputenc}

\begin{document}

\title{Infrared chiral anomaly at finite temperature}

\author{Ashok Das$^{a,b}$ and J. Frenkel$^{c}$\footnote{$\ $ e-mail: das@pas.rochester.edu,  jfrenkel@fma.if.usp.br}}
\affiliation{$^a$ Department of Physics and Astronomy, University of Rochester, Rochester, NY 14627-0171, USA}
\affiliation{$^b$ Saha Institute of Nuclear Physics, 1/AF Bidhannagar, Calcutta 700064, India}
\affiliation{$^{c}$ Instituto de Física, Universidade de São Paulo, 05508-090, São Paulo, SP, BRAZIL}

\begin{abstract}
We study the Schwinger model at finite temperature and show that a temperature dependent chiral anomaly may arise from the long distance behavior of the electric field. At high temperature this anomaly depends linearly on the temperature $T$ and is present not only in the two point function, but also in all even point amplitudes.
\end{abstract}


\maketitle

\centerline{(Dedicated to the memory of Olivier Espinosa.)}
\bigskip

Quantum field theories are well known to have ultraviolet divergences which arise from the short distance behavior of the theory. These ultraviolet divergences need to be regularized (so that the theory makes sense) and one normally tries to choose a regularization scheme which would preserve the underlying symmetries of the quantum field theory. This requirement of choosing a regularization preserving the symmetry becomes particularly crucial when one is dealing with a gauge theory (with a local symmetry) in order to avoid possible inconsistencies in the theory. However, when there are several distinct symmetries present in a theory, it may not always be possible to find a regularization which respects all the symmetries of the theory in which case one tries to choose a regularization which respects the gauge symmetries of the theory. In this case, the regularization may not respect some of the global symmetries of the theory and may lead to nontrivial quantum corrections to the conservation laws for the currents associated with them. A deviation from the classical conservation law is normally called an anomaly and the chiral anomaly \cite{schwinger,adler,jackiw,adler-bardeen,bardeen} is one of the well studied examples of such  phenomena. It is also well known that while ultraviolet divergences are present in a quantum field theory at zero temperature, temperature dependent parts of amplitudes are ultraviolet  finite and  do not require any additional counterterm at finite temperature. Indeed since we do not need to regularize the temperature dependent parts of the amplitudes, it follows that there will not be any temperature dependent correction to anomalies and this has been explicitly checked in the case of the chiral anomaly (in various theories) \cite{dolan,itoyama,karev,das}. 

Although there is no ultraviolet divergence in a quantum field theory at finite temperature, the infrared behavior is more pronounced and thermal infrared divergences are known to be severe. Therefore, it is natural to ask if the infrared behavior of a quantum field theory at finite temperature can lead to contributions for the chiral anomaly. We note that if this were to happen, it would clearly depend on the large distance behavior of a quantum field theory. In fact, since quite often one studies quantum fields in a constant background field (constant electric/magnetic field), such an investigation is quite meaningful. In this letter we undertake a systematic investigation of this question for general electromagnetic background fields (coupled to massless fermions in $1+1$ dimensions) which do not vanish asymptotically and determine which  backgrounds can contribute to the chiral anomaly at finite temperature. In $1+1$ dimensional massless QED (the Schwinger model \cite{schwinger2}) the chiral anomaly at zero temperature  is normally associated with the two point function which is the only diagram in the theory with an ultraviolet divergence. In contrast, when the chiral anomaly has its origin in the infrared behavior of the theory, we show that higher point amplitudes can also become anomalous and we determine the complete temperature dependent anomaly functional (in the leading order at high temperature) for the Schwinger model. We work in the real time formalism of finite temperature known as the closed time path formalism due to Schwinger \cite{das,schwinger1} and restrict our discussion, for simplicity, to the ``$+$" branch of the contour (see \cite{das} for details). 

The Schwinger model \cite{schwinger2} is described by (we are neglecting the Maxwell term for the photon) the $1+1$ dimensional  Lagrangian density
\begin{equation}
{\cal L} = \overline{\psi} \gamma^{\mu} \left(i\partial_{\mu} - e A_{\mu}\right)\psi,\label{L}
\end{equation}
where $\mu=0,1$ and the Dirac matrices are identified with the Pauli matrices as $\gamma^{0}=\sigma_{1}, \gamma^{1}=-i\sigma_{2}, \gamma_{5}=\gamma^{0}\gamma^{1}=\sigma_{3}$. The fermions are massless in this theory and let us first show that if they had a mass, there would not be any temperature dependent contribution to the chiral anomaly even when the background fields do not vanish asymptotically. Therefore, the phenomenon that we are describing has a genuine infrared origin.

\begin{figure}[ht!]
\includegraphics[scale=.7]{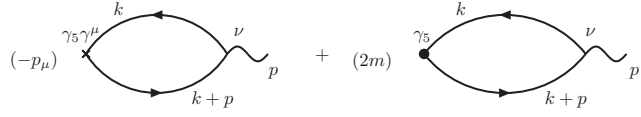}
\caption{The two diagrams contributing to \eqref{massiveWI} at the two point level.}
\label{1}
\end{figure}

To see that the massive theory does not generate any temperature dependent contribution to the anomaly \cite{frenkel} (even when the background field does not vanish asymptotically), let us note that the chiral current in the massive Schwinger model satisfies the tree level identity
\begin{equation}
\partial_{\mu} J^{\mu}_{5} (x) + 2im J_{5}(x) = 0,\label{massiveWI}
\end{equation}
where $J^{\mu}_{5} = \overline{\psi}\gamma_{5}\gamma^{\mu}\psi, J_{5} = \overline{\psi}\gamma_{5}\psi$ and $m$ denotes the mass of the fermions. At one loop the contribution to this relation at the two point level comes from the two diagrams in Fig. \ref{1}.

In the closed time path formalism \cite{das,schwinger1}, the temperature dependent contribution from these diagrams (say, on the $++$ branch)  is obtained from
\begin{align}
\lefteqn{\epsilon^{\mu\nu}A_{\nu}(p)\!\!\int \!\mathrm{d}^{2}k\!\left[(k^{2}-m^{2})(k+p)_{\mu} - ((k+p)^{2}-m^{2}) k_{\mu}\right]}\notag\\
&\qquad\qquad\times\left(\!\frac{i}{(k+p)^{2}-m^{2}+i\epsilon} - \pi  n_{\rm F} (|k^{0}+p^{0}|) \delta ((k+p)^{2}-m^{2})\right)  n_{\rm F} (|k^{0}|)  \delta(k^{2}-m^{2}),\label{anomaly}
\end{align}
where $\epsilon^{\mu\nu}$ denotes the two dimensional Levi-Civita tensor and we use the convention $\epsilon^{01}=1$. The single delta function terms in \eqref{anomaly} vanish because of delta function constraints as well as anti-symmetry. The nontrivial contribution, if present, can arise only from the two delta function terms in \eqref{anomaly}. The contributions from the two delta function terms would clearly vanish if the two delta function constraints truly restrict $k_{0}$ and $k_{1}$ because of the structure of the terms in the square brackets in \eqref{anomaly}. This is, in fact, true in the massive theory and there is no temperature dependent contribution to the anomaly. However, to appreciate why the contribution may not vanish in the massless theory (Schwinger model), let us analyze the constraints following from the product of two delta functions.

For example, in the massive theory we note that the two delta function constraints can be satisfied only if $p^{2}\neq 0$ which follows from the kinematics of the diagrams since the loop involves propagation of massive particles. In fact, the diagrams contribute only if $1- \frac{4m^{2}}{p^{2}} \geq 0$ which is the origin of the two branch cuts $p^{2}\geq 4m^{2}$ and $p^{2}\leq 0$ in the two point function at finite temperature \cite{das,weldon}. In such a case, the product of the two delta functions can be factorized as
\begin{align}
\lefteqn{\delta(k^{2}-m^{2})\delta ((k+p)^{2} - m^{2})
 = \frac{\theta (\omega_{+}(p_{0}))}{2\omega_{+}(p_{0})}\delta (k_{0}-\omega_{+} (p_{0}))\left[\frac{1}{|J_{+,+} (p_{0})|}\delta (k_{1}-X_{+} (p_{0})) + \frac{1}{|J_{+,-} (p_{0})|} \delta (k_{1}+ X_{+} (p_{0}))\right]}\notag\\
&+\ \frac{\theta (\omega_{-}(p_{0}))}{2\omega_{-}(p_{0})}\delta (k_{0}-\omega_{-}(p_{0}))\left[\frac{1}{|J_{-,+} (p_{0})|}\delta (k_{1}-X_{-} (p_{0})) + \frac{1}{|J_{-,-} (p_{0})|} \delta (k_{1}+ X_{-} (p_{0}))\right] + (k_{0},p_{0}\rightarrow -k_{0}, -p_{0}),\label{factorizationm}
\end{align}
where we have identified 
\begin{align}
\omega_{\pm} (p_{0}) & = \frac{1}{2}\left(-p_{0} \pm |p_{1}|\sqrt{1 - \frac{4m^{2}}{p^{2}}}\right),& & X_{\pm} (p_{0})  = \sqrt{\omega_{\pm}^{2} (p_{0}) - m^{2}},\notag\\
J_{+,\pm} (p_{0}) & = \frac{\mp X_{+}(p_{0})p^{2} - 2m^{2}p_{1}}{\omega_{+}^{2} (p_{0})},&  
& J_{-,\pm} (p_{0})  = \frac{\mp X_{-}(p_{0})p^{2} - 2m^{2}p_{1}}{\omega_{-}^{2} (p_{0})}.
\end{align}
The product of the two delta functions clearly constrains the integration variables $k_{0},k_{1}$ and in such a case the terms in the square bracket in \eqref{anomaly} make the integral vanish independent of the behavior of the background field $A_{\nu} (p)$.

In a massless theory ($m=0$), on the other hand, the momenta naturally separate into light-cone components and the product of two delta functions in this case factorize as
\begin{align}
\delta(k^{2})\delta ((k+p)^{2}) & = \frac{1}{|k_{-}||p_{+}|} \delta (k_{+}) \delta (k_{-}+p_{-}) + \frac{1}{|k_{+}||p_{-}|} \delta (k_{-}) \delta (k_{+}+p_{+})\notag\\
&\quad + \frac{1}{|k_{-}||k_{-}+p_{-}|}\delta(k_{+})\delta(p_{+}) + \frac{1}{|k_{+}||k_{+}+p_{+}|} \delta(k_{-})\delta(p_{-}).\label{factorization}
\end{align}
Comparing this with \eqref{factorizationm}, we note that the first two terms in \eqref{factorization} constrain the variables of integration completely and lead to vanishing contribution as in the massive theory. However, the last two terms in \eqref{factorization} constrain only partially the variables of integration and as a result the second term in the square bracket leads to a nontrivial contribution for \eqref{anomaly} of the form 
\begin{equation}
\left(p_{+}\delta(p_{+})A_{-} (p) - p_{-}\delta(p_{-})A_{+} (p)\right)I_{2},\label{anomaly1}
\end{equation}
where as discussed in \cite{adilson,frenkel1,frenkel2} ($\epsilon (x)$ denotes the alternating step function)
\begin{equation}
I_{2} = \frac{(2ie\pi)^{2}}{2} \int \frac{dk_{1}}{(2\pi)^{2}} \epsilon(k_{1}) \epsilon(k_{1}+p_{1})\left(n_{\rm F} (|k_{1}|) + n_{\rm F} (|k_{1}+p_{1}|) - 2 n_{\rm F} (|k_{1}|)n_{\rm F} (|k_{1}+p_{1}|)\right).\label{I2}
\end{equation}
Normally, for background fields vanishing asymptotically, \eqref{anomaly1} vanishes leading to the conclusion that there is no temperature dependent contribution to the anomaly. However, we note that if $A_{+} (p)\sim \frac{1}{p_{-}}$ or $A_{-} (p) \sim \frac{1}{p_{+}}$ asymptotically (for, example), then \eqref{anomaly1} would not vanish and give rise to a temperature dependent contribution to the chiral anomaly. This is the infrared anomaly that we have mentioned earlier and which we will now discuss in more detail.

Let us note that the integral \eqref{I2} cannot be evaluated in closed form in general. However, in the leading order in the hard thermal loop expansion, the dependence on the external momenta in $I_{2}$ can be neglected and we can determine
\begin{equation}
I_{2}  = \frac{(2ie\pi)^{2}}{2} \int_{0}^{\infty} \frac{dk_{1}}{(2\pi)^{2}} \left[1 - (1 - 2 n_{\rm F} (k_{1}))^{2}\right] = (ie)^{2} T \int_{0}^{\infty} dx\,\left(1-\tanh^{2} x\right) = (ie)^{2} T\int_{0}^{1} d(\tanh x) = (ie)^{2} T.\label{I2_1}
\end{equation}
With this, \eqref{anomaly1} can be Fourier transformed to the coordinate space (in a regularized manner so as to take care of products of singular functions) leading to the temperature dependent correction to the anomaly at the two point level (in the massless theory) as
\begin{equation}
\partial_{\mu} J_{5}^{\mu (\beta)} (x) = -\frac{(ie)^{2} T}{\pi}\!\left[\int\!\! dy^{+} {\rm sgn} (x^+ - y^+) \!\left(E (y^+, \infty) - E (y^+, -\infty)\right) + \int\!\! dy^- {\rm sgn} (x^- - y^-)\!\left(E (\infty, y^-) - E (-\infty, y^-)\right)\!\right]\!,\label{2point}
\end{equation}
where $E(x^+,x^-)$ denotes the electric field and sgn$(x-y)$ represents the sign function. Equation \eqref{2point} shows that if the electric field vanishes asymptotically, there is no temperature dependent contribution to the anomaly. In fact, electric fields satisfying
\begin{equation}
E (x^{+},x^{-}) = E (x^{+},-x^{-}) = E (-x^{+},x^{-}),\label{condition}
\end{equation} 
do not contribute to \eqref{2point} even if they do not vanish (go to a constant) asymptotically. As a result, a constant background electric field does not contribute to a temperature dependent anomaly in \eqref{2point}. On the other hand, an electric field which does not satisfy \eqref{condition}, for example,
\begin{equation}
E (x^{+},x^{-}) = E_{1}\,{\rm sgn} (x^{+}) \delta (x^{-}),\quad {\rm or}\quad E(x^{+},x^{-}) = E_{2}\,{\rm sgn} (x^{-}) \delta (x^{+}),\label{efield}
\end{equation}
would lead to a temperature dependent anomaly at the two point level which has the form, say for the first case in \eqref{efield},
\begin{equation}
\partial_{\mu}J^{\mu (\beta)}_{5} (x) = - \frac{2E_{1}\,(ie)^{2}\, T}{\pi}\, {\rm sgn} (x^{-}).\label{2pointanomaly}
\end{equation}
We call such an anomaly an infrared anomaly since it arises from the long distance behavior of the electric field as opposed to the conventional anomaly which has its origin in the ultraviolet behavior. It is also worth noting that the anomaly in \eqref{2point} results because of products of delta functions (see \eqref{factorization}) that arise at finite temperature and which are not present at zero temperature. Therefore, the infrared anomaly is not present at zero temperature.

The conventional anomaly in the Schwinger model is present only at the two point level since the higher point functions are ultraviolet finite. However, since the infrared anomaly is not associated with ultraviolet divergence, in principle, it may manifest in higher point functions as well. We will show that this is indeed true and calculate the complete anomaly functional. We note that the complete temperature dependent one loop effective action for the Schwinger model has been determined in \cite{adilson,frenkel1,frenkel2} and has the leading behavior at high temperature given by
\begin{equation}
\Gamma_{\rm eff}^{(\beta)} = \sum_{n} \Gamma_{2n}^{(\beta)} = \sum_{n} \frac{1}{2n!}\int \prod_{i=1}^{2n}\left(\frac{dp_{i}}{(2\pi)^{2}} (\bar{u}\cdot A) (p_{i})\right) \delta^{2} (p_{1}+\cdots + p_{2n}) \left(\prod_{j=1}^{2n-1} \delta (\omega_{j}-p_{j}) +\prod_{j=1}^{2n-1} \delta (\omega_{j}+p_{j})\right)\, I_{2n}^{(\beta)}.\label{effaction}
\end{equation}
Here the transverse velocity of the heat bath is defined as (see \cite{frenkel} for notations)
\begin{equation}
\bar{u}^{\mu}(p) = u^{\mu} - \frac{\omega}{\bar{p}} \epsilon^{\mu\nu} u_{\nu},\quad \omega = u\cdot p,\quad \bar{p} = \epsilon^{\mu\nu} p_{\mu}u_{\nu},
\end{equation}
and the temperature dependent factor $I_{2n}^{(\beta)}$ in general has a complicated form  (see \eqref{I2}). However, in the hard thermal loop limit where the dependence on the external momenta can be neglected it has the simpler form (see also \eqref{I2_1})
\begin{align}
I_{2n}^{(\beta)} & = \frac{(2\pi ie)^{2n} (2n-1)!}{2^{2n-1}}\int\limits_{0}^{\infty} \frac{dk_{1}}{(2\pi)^{2}} \left[1 - (1-2n_{\rm F} (k_{1}))^{2n}\right] = \pi^{2(n-1)} (2n-1)! (ie)^{2n} T \int\limits_{0}^{1} d(\tanh x)\sum_{m=1}^{n} \tanh^{2(m-1)}x\notag\\
& = \pi^{2(n-1)} (2n-1)! (ie)^{2n}\,C_{2n}\, T,\qquad C_{2n} = \sum_{m=1}^{n} \frac{1}{2m-1}.\label{I2n_1}
\end{align}
With this the leading behavior of the effective action simplifies considerably. Furthermore, we would work in the rest frame of the heat bath for simplicity where $u^{\mu} = (1,0)$ and $\bar{u}^{\mu} = \frac{\epsilon^{\mu\nu} p_{\nu}}{p_{1}}$ which leads to $\bar{u}\cdot A (p) = - \frac{\tilde{E} (p)}{p_{1}}$ where $\tilde{E}(p)$ denotes the electric field in the momentum space.

Given the temperature dependent effective action in $1+1$ dimensions \eqref{effaction} we can obtain the temperature dependent vector current as well as the axial vector current by taking the functional derivative with respect to the background gauge field. For example,
\begin{align}
J_{5}^{\mu (\beta)} (P) & = \sum_{n=1} J_{5, 2n-1}^{\mu (\beta)} (P) = \epsilon^{\mu\nu}\, \frac{\delta\Gamma_{\rm eff}^{(\beta)}}{\delta A_{\nu} (-P)}\notag\\
& = \sum_{n=1} \frac{(ie)^{2n}}{(2\pi)^{2}}\int \prod_{i=1}^{2n-1}\left(d^{2}p_{i}\left(-\frac{\tilde{E} (p_{i})}{p_{i, 1}}\right)\right)\frac{P^{\mu}}{P_{1}} \delta^{2} (P-\sum_{i=1}^{2n-1} p_{i}) \left(\prod_{i=1}^{2n-1} \delta(p_{i,+}) + \prod_{i=1}^{2n-1} \delta(p_{i,-})\right)C_{2n}\, T.
\end{align}
The anomaly can now be calculated for any value of $n$ and has the form
\begin{align}
A_{2n-1}^{(\beta)} (P)  = P_{\mu} J_{5, 2n-1}^{\mu (\beta)} (P) & =  - \frac{(2ie)^{2n} C_{2n}}{(2\pi)^{2}}\bigg[\int d^{2}p_{2n-1} \left(\prod_{j=1}^{2(n-1)} \left(d^{2}p_{j} \frac{\tilde{E} (p_{j})}{p_{j, 1}} \delta (p_{j, +})\right)\right)\notag\\
& \quad \times p_{2n-1,+} \delta (p_{2n-1,+}) \frac{\tilde{E} (p_{2n-1})}{p_{2n-1,-}}\delta (P-\sum_{i=1}^{2n-1} p_{i}) + p_{i,\pm}\rightarrow p_{i, \mp} + {\rm permutations}\bigg]T.
\end{align}
Once again we see that for fields that vanish asymptotically this temperature dependent anomaly vanishes, but for electric fields with a singular behavior $\tilde{E} (p) \sim \frac{1}{p_{\pm}}$, this leads to a finite contribution. In fact, Fourier transforming this to the coordinate space we obtain (the product of singular functions needs to be handled in a regularized manner)
\begin{equation}
A_{2n-1}^{(\beta)} (x) = - (ie)^{2n} \pi^{2n-3} (2n-1) C_{2n}\left[\left(I (x^{+})\right)^{2(n-1)} J (x^{+}) + \left(I (x^{-})\right)^{2(n-1)} J (x^{-})\right] T,\label{2npoint}
\end{equation}
where we have defined $I (x^{\pm}) = \int d^{2}y\,{\rm sgn} (x^{\pm} - y^{\pm}) E (y)$ as well as the boundary terms 
\begin{equation}
J (x^{+}) = \int dy^{+}\,{\rm sgn} (x^{+}-y^{+}) \left(E(y^{+},\infty) - E (y^{+},-\infty)\right),\ J(x^{-}) = \int dy^{-}\,{\rm sgn} (x^{-}-y^{-}) \left(E(\infty,y^{-}) - E(-\infty, y^{-})\right).
\end{equation}
This vanishes for a constant field configuration. However, for a field configuration of the type in \eqref{efield}, for example for the first configuration, it leads to 
\begin{equation}
A_{2n-1}^{(\beta)} (x) = - (2E_{1})^{2n-1} (ie)^{2n} \pi^{2n-3} (2n-1) C_{2n}\,|x^{+}|^{2(n-1)}\,{\rm sgn}(x^{-})\, T.
\end{equation}
Equation \eqref{2npoint} shows that the temperature dependent infrared anomaly is present at every even point level and the complete temperature dependent anomaly functional is given by
\begin{equation}
A^{(\beta)} = \sum_{n=1} A_{2n-1}^{(\beta)},
\end{equation}
and depends linearly on the temperature in the high temperature limit.

Even though we have discussed the temperature dependent infrared anomaly in the $1+1$ dimensional Schwinger model, such a phenomenon is quite likely to arise in $3+1$ dimensional gauge theories as well which are expected to have strong infrared behavior at higher loops. It is interesting to understand the connection of this infrared anomaly with the index theorem. An equally interesting and related question to ask is since such an anomaly is not related to the ultraviolet behavior of the theory, whether the Adler-Bardeen theorem \cite{adler-bardeen} continues to hold or whether there can be higher loop corrections to this anomaly. These are some of the questions presently under study.

\bigskip

\noindent{\bf Acknowledgments}
\medskip

This work was supported in part  by US DOE Grant number DE-FG 02-91ER40685,  by CNPq and FAPESP (Brazil).

\end{document}